\documentclass{article}
\begin{document}
\begin{center}
{\large \bf Three qubit GHZ correlations and generalised Bell experiments} \\

Dagomir Kaszlikowski$^1$ and Marek \.Zukowski$^3$\\
{\protect{$^1$ Faculty of Science, National University of Singapore,\\  Singapore 117542\\
$^2$Instytut Fizyki Do\'swiadzalnej,\\
Uniwersytet Gda\'nski, PL-80-952 Gda\'nsk, Poland\\
$^3$Instytut Fizyki Teoretycznej i Astrofizyki,\\
Uniwersytet Gda\'nski, PL-80-952 Gda\'nsk, Poland
}}
\end{center}

\abstract{We present a brief historical introduction to the topic of Bell's theorem. 
Next we present the  surprising features  of the 
three particle Greenberger-Horne-Zeilinger (GHZ)  states.  
Finally we  shall  present a  method  of analysis of the  GHZ  correlations, which is based 
on a numerical approach, which is effectively equivalent to the  full set of Bell inequalities 
for correlation functions for the given problem.
The aim of our numerical approach is to answer the following question. Do additional possible local 
settings lead for the GHZ states to more pronounced violation of local realism (measured by the resistance 
of the quantum nature of the  correlations with respect ``white'' noise admixtures)?
}

\section{Introduction: the early history of the problem}

In the introduction part of this  paper we would like to give a picture to the readers, especially the
 young ones, of the  years when the term quantum information was not yet invented, however some basic 
research, that later gave birth to this new branch of physics, already began.
  
 In 1964 Bell (1964) demonstrated, that no local and realistic (that is classical relativistic) theory
 could ever agree with all  predictions of quantum
   mechanics. His theorem showed that the idea of Einstein,
   Podolsky, and Rosen (1935) (EPR) of completing quantum mechanics, so that that the 
resulting theory would be deterministic, is impossible. The theorem of Bell
 raises the following profound 
   question:  can one model natural phenomena with a local relativistic theory? 
Moreover, the theorem provided a  blueprint for an experimental test of this
problem. 

The consequences of the Bell theorem are so dramatic, so that 
experiments were needed. 
But there were two problems. the first one was that the original 
Bell's inequality required from the quantum two-particle system 
to possess 
perfect correlations. This is possible in theory, and indeed such entangled 
two-particle states exits in the Hilbert Space, however in the laboratory this impossible.
 Simply in every experiment some noise is inevitable.
The second problem was to find a source of entangled states 
that could give rise to observable quantum effects, that could be used in a Bell test. 
Both problems were solved in the trail-blazing paper by 
Clauser, Horne, Shimony and Holt (1969), the famous CHSH paper.
A new type of Bell inequality was proposed, which did not require
the the prefect correlations. It gives a bound of correlations
describable by local realistic theories. But this is not all. 
The authors noticed that the two photon cascades in Calcium result in emissions of 
pairs of 
photons with entangled polarisations. I.e., a source for 
the pairs of entangled particles needed for a test of Bell's inequality (now 
rather the CHSH one) was found, and the two photon emissions, already an interesting phenomenon,
were shown to
be a very exotic effect in this case. Had this effect been known to Einstein and his colleagues,
most probably the EPR paper would have been completely different.
Since now a lot of experimental  and theoretical physics is devoted to studying
entangled  states (usually in the form of entangled polarisations), more, even 
a whole new branch of physics emerged (quantum information) which tries  to 
understand and exploit as a resource entanglement, this pinpointing by CHSH
of the first controllable source of entanglement deserves to be called a great discovery. 
Yes, earlier entangled states were present both is theory and experiment but 
their drastically non-classical properties were never observed directly.
The new source enabled precisely this.

In few years time the actual 
experiment was performed by  Freedman and Clauser (1972). To the amazement of many, Bell's 
inequalities were violated by a natural  phenomenon observed in the lab.
The prediction of the existence of entangled states of light was experimental confirmed. 

Another important step was  the paper by Clauser and Horne
(1974) in which they derived yet another Bell type inequality, now 
customarily called the CH one. This inequality has several very important features. 
It is  testable (like the CHSH one), it implies the CHSH one 
(but the CHSH does not imply the CH one), and  is perfect for the analysis of the 
threshold parameters required for a ``loop-hole free'' Bell test.

The dramatic consequences of Bell's theorem, and the 
falsification of local realism in the experiments of Clauser, 
caused a reaction in camp of researchers who were sceptical  
about the universal 
validity or completeness of quantum theory. This reaction resulted 
in many papers in which the above mentioned ``loop-holes'', 
i.e. imperfections of the Freedman-Clauser test of Bell inequalities, 
were studied, and which according to their authors could invalidate the 
experiment as a falsification of local realism.

Aspect et al (1982) designed 
and performed  Bell experiments aimed  a closing on of the loopholes. The same Calcium cascade 
was the source, but 
much more effective pumping (by lasers) was used, and therefore the 
statistics was now much better. But
especially important was the
   experiment in which the polarisations to be measured at 
two far away detection stations were set effectively
during the flight of the photons. This guaranteed that the measurement
setting at side A was absolutely unknown at side B at the moment of 
the detection of the photon (and vice versa). Thus any spooky theory that
 could ``explain'' the quantum correlations via a local and 
realistic model, employing the ``loop-hole'' of the Clauser experiment 
(namely fixed  polariser settings 
throughout the each experimental run) was closed. In this way it was for the
first time experimentally established that realistic theories of nature must be necessarily 
non-local. This was the most important loophole to be closed. Other loop-holes are associated 
only with imperfections
in the measuring devices (like detector efficiency). However since the quantum 
predictions are so well reproduced already in an imperfect experiment, why should we expect
some deviations in a more precise one? 

To summarise,  the above  experimental tests of 
local realism falsified this idea (so cherished  by e.g.  Einstein), and in that way solved 
the Einstein-Bohr debate (almost) definitely in favour of Bohr. 
The existence in nature of entangled  states was 
experimentally confirmed.
Much later, it was experimentally proved
 that entanglement can be utilised directly in quantum cryptography (the protocol of Ekert, 1991). 
Quantum cryptography is now already within the realm of applied physics. Entanglement is essential in 
the process of quantum teleportation, and in various quantum communication and quantum information schemes.
The very topic of entanglement leads to such surprises like the ultra-non-classical 
Greenbeger-Horne-Zeilinger (1989) correlations. But in the beginning of all that were the ideas of Bell, 
and the early  experiments. They showed with new 
strength, how strange is the quantum world, and that this  strangeness can be experimentally observed.
Now  we  begin to benefit from that. 

\section{Summary}

Further down in the paper we shall present a brief  introduction to the  Bell theorem, 
which will lead us to the simplest Bell inequality (as far as 
the derivation is concerned), which is the aforementioned CHSH  one.
Next we shall give a brief of introduction to the  surprising features  of the 
three particle Greenberger-Horne-Zeilinger (1989) (GHZ)  states. 
Finally we  shall  present a  new method  of analysis of the  GHZ  correlations, which is based 
on a numerical approach, which is effectively equivalent to the  full set of Bell inequalities 
for correlation functions for the given problem.\footnote{The set of Bell inequalities is  ``full''
when it constitutes the sufficient and  necessary  condition for the existence of local realistic model 
for the given process, for the given number  of local settings for each of the observers.} 
Such a set is  known in the case of three qubits in the 
case of experiments  involving two alternative  settings for each observer 
(Weinfurter and \.Zukowski, 2001, Werner and Wolf, 2001, \.Zukowski and  Brukner, 2002). 
In this case  the full set can be expressed in the form 
of a single generalised inequality, and therefore it is easy to analyse. The full set of inequalities for
three settings per observer can be in principle found using the method presented by Pitovsky and  Svozil (2001).
However, one can expect an enormous  number of them. 

The aim of our numerical approach is to answer the following question. Do additional possible local 
settings lead, in the case of three qubit
GHZ states to more pronounced violations of local realism (measured 
by the resistance of the quantum nature of the  correlations to ``white'' noise admixtures)?

\section{Bell theorem}
Before the advent of Bell (1964) theorem, despite Einstein's
doubts, the question of the existence of a more detailed
description of individual events in the micro-world, than the
probabilistic one provided by quantum mechanics,  was treated as
interesting, however not falsifiable, and therefore  as irrelevant
as the question of `how many angels fit on the  tip  of the
needle''.  In early sixties Bell (1966)\footnote{Bell
(1966) was written before Bell (1964). } conjectured, that if
there is any conflict between quantum mechanics and the {\it
realistic theories}\footnote{Realism, the cornerstone of classical
physics, is  a view that any physical system (i.e. also a
subsystem of a compound system) carries full information
(deterministic or probabilistic) on results of {\it all} possible
experiments that can be performed upon it.}, it may be confined to
{\it local}\footnote{A theory is local if it assumes that
information cannot travel faster than light.}
 versions
of such theories. This led him to formulate his famous theorem, of
profound scientific and philosophical consequences.

\subsection{Bell inequality}

Consider pairs of particles (say, photons) simultaneously  emitted
in well defined opposite directions. After some time the photons
arrive  at two  very distant measuring devices A and B operated by
two characters: Alice and Bob.
 Their apparatuses,
are have a knob which specifies,  which dichotomic (i.e.,
two-valued, yes-no, 0-1, one bit) observable they actually measure
\footnote{E.g., for a device consisting of a polarising
beam-splitter and two detectors behind its outputs, this knob would
specify the orientation of the polariser; if the device is a
Mach-Zehnder interferometer (plus two detectors at the two exits)
the knob would set the phase shift, etc. The photon may be
registered only behind one  or  the other output ports of such
devices.}. One can assign to the two possible results the numbers
$+1$ (for yes, bit value one) and $-1$ (for no, bit value
nil)\footnote{We assume perfect situation in which the detectors
never fail to register a photon.}.
 Alice and Bob are at any time (also in a `delayed choice'
mode, after an emission) {\it both free} to independently choose the observables (knob
settings) that they want to measure. Their choice is absolutely independent
of the workings of the source, and can be done at any time.

Let us assume that the each photon pair carries full information
(deterministic or probabilistic)  on the values of the results of
all possible experiments that can be performed on
it\footnote{Note, that in the present discussion,  only this idea
openly goes beyond `what is speakable' in quantum mechanics.}
({\it realism}). Also, by {\it locality}, choices made by them
which are simultaneous in certain reference frame cannot influence
each other (in Alice's region of space-time, which contains the
measurement event, there is no information whatsoever available on
Bob´s choice, and {\em vice versa}); the choice made on one side
cannot influence the results on the other side.

For simplicity, assume that
 Alice, chooses to measure  either observable $\hat{A_1}$
or $\hat{A_2}$, and Bob either $\hat{B_1}$ or $\hat{B_2}$.
 Let us denote the hypothetical results that they may get for the $j$-th pair by
$A_1^j$ and  $A_2^j$, for Alice´s two possible choices, and
$B_1^j$ and $B_1^j$, for Bob´s. The numerical values of these
results ($+1$ or $-1$) are defined by the two eigenvalues of the
observables. Since, always either $|B_1^j -B_2^j|=2$ and $|B_1^j
+B_2^j|=0$, or $|B_1^j -B_2^j|=0$ and $|B_1^j +B_2^j|=2$,
 the following
relation holds

\begin{eqnarray}
&A_1^j B_1^j + A_1^jB_2^j+A_2^jB_1^j-A_2^jB_2^j& \nonumber \\
&=A_1^j \big(B_1^j +B_2^j\big)+A_2^j\big(B_1^j-B_2^j\big)=\pm2.&
\label{BELLEQ}
\end{eqnarray}

  Imagine now that  $N$
pairs of photons to are emitted, pair by pair ($N$ is
 sufficiently large, $\sqrt{1/N}\ll 1$).
 The average value of the products of the local
values for a joint test (often called the Bell correlation
function) during which, for all photon pairs, only one pair of
observables, say $\hat{A_n}$ and $\hat{B_m}$, is chosen by the
local observers is given by
\begin{equation}
E(A_n,B_m)=\frac{1}{N}\sum_{j=1}^{j=N}A_n^jB_m^j,
\end{equation}
where $n=1,2$ and $m=1,2$.  The
 relation  implies that for the four {\it possible}
 choices of pairs of observables the following ``Bell" inequality
 must be satisfied
\begin{eqnarray}
&-2\leq E(A_1,B_1)+E(A_1,B_2)&\\ \nonumber &+E(A_2,B_1)-
E(A_2,B_2)\leq2,&\\ \label{BELLINEQ}
\end{eqnarray}
(Clauser, Horne, Shimony and Holt, 1969). In the actual experiment
only in part of the cases (say, approximately $1/4$-th) the given
pair of observables would be measured, however if $N$ is very
large, the correlation function function obtained on a randomly
pre-selected sub-ensemble\footnote{The submersible is selected by
the choice of observables made by Alice and Bob {\em before} the
actual measurements.} of emissions cannot differ too much from the
one that would have been obtained for the full
ensemble\footnote{With $N\rightarrow\infty$ the difference must
approach zero,
 otherwise we would suspect that these two
magnitudes pertain to two different physical processes (i.e., a
systematic error must be involved).}
 Therefore for the values of the {\it actually} chosen measurements the
 inequality  also must hold\footnote{The
presented Bell-type argument avoids  any explicit introduction of
hidden variables. All {\it local} hidden variable theories satisfy
Bell inequalities.}.

\subsection{Bell theorem without inequalities: three entangled particles or more}

If there are $N>2$ maximally entangled quantum systems (qubits),
measurements on which are performed in $N$ spatially separated
regions by $N$ independent observers, the correlations obtained in
such experiment violate bounds imposed by local realism much
stronger than in the case of two particles.  Proofs of such
violations can be limited to the 
experiments for which the perfect EPR-type correlations occur.
That is, the EPR notion of elements of reality becomes
self-contradictory.

As the simplest example, take a GHZ (Greenberger, Horne,
Zeilinger, 1989) state of $N=3$ particles (fig.1):
\begin{equation}
|\psi(3)>
= \frac{1}{\sqrt{2}}\left(|a\rangle |b\rangle |c\rangle
+|a'\rangle |b'\rangle |c'\rangle\right)
\label{1}
\end{equation}
where $\langle x|x'\rangle=0$ ($x=a,b,c$, and kets denoted by one
letter pertain to one of the particles). The  observers, Alice,
Bob and Cecil  measure the observables: $\hat{A}(\phi_A)$,
$\hat{B}(\phi_B)$, $\hat{C}(\phi_C)$, defined by

\begin{equation}
\hat{X}(\phi_X)
 = |+,\phi_X\rangle  \langle+,\phi_X |
- |-,\phi_X\rangle  \langle-,\phi_X |
\label{2}
\end{equation}
and
\begin{equation}
 |\pm,\phi_X\rangle
= \frac{1}{\sqrt{2}}\left(\pm i|x'\rangle + \exp{(i\phi_X )}|x\rangle\right)
.
\label{3}
\end{equation}
where $X=A,B, C$.
The quantum prediction for the expectation value
of the product of the three
local observables is given by
\begin{eqnarray}
&E(\phi_A ,\phi_B ,\phi_c)
=  <\psi|\hat{A}(\phi_A)\hat{B}(\phi_B)\hat{C}(\phi_C)
|\psi>&\nonumber\\
& =  sin(\phi_A +\phi_B+\phi_c). &\\
\nonumber
\end{eqnarray}
Therefore, if $\phi_A +\phi_B+\phi_c=\pi/2+k\pi$, quantum
mechanics predicts perfect correlations. E.g., for $\phi_A=\pi/2$,
$\phi_B=0$ and $\phi_c=0$, whatever may be the results of local
measurements of the observables, for say the particles belonging
to the $i$-th triple represented by the quantum state
$|\psi(3)\rangle$, they have to satisfy
\begin{equation}
A^i(\pi/2) B^i(0) C^i(0) = 1, \label{4}
\end{equation}
where $X^i(\phi)$, $X=A,B$ or $C$ is the value  of a local
measurement of the observable $\hat{X}(\phi)$ that {\it would have
been} obtained for the $i$-th particle triple if the setting of
the measuring device  is  $\phi$. By locality $X^i(\phi)$ depends
solely on the local parameter. The eq. (\ref{4}) indicates that we
can predict with certainty the result of measuring the observable
pertaining to one of the particles (say $c$) by choosing to
measure suitable observables for the other two. Hence the value
$X^i(\phi)$ are EPR elements of reality.

However, if the local apparatus settings are different one {\it
would have had}, e.g.
\begin{eqnarray}
&A^i(0) B^i(0) C^i(\pi/2) = 1,&\\ \label{5}
 &A^i(0) B^i(\pi/2)
C^i(0) = 1,&\\ \label{6}
 &A^i(\pi/2) B^i(\pi/2) C^i(\pi/2) = -1.&
\label{7}
\end{eqnarray}
 Since $X^i(\phi)\pm1$, if one multiples side by
side  the eqs (\ref{4}-\ref{6}), the result is
\begin{equation}
 A^i(\pi/2) B^i(\pi/2) C^i(\pi/2) = +1,
\label{8}
\end{equation}
which contradicts (\ref{7}).
 Thus the EPR elements of reality program
breaks down. We have a ``Bell theorem without inequalities"
(Greenberger, Horne and Zeilinger, 1989).

\section{More than two settings per each observer }

The beautiful argument of the  GHZ  paper cannot be directly applied to experimental
results. This just like in the case of the original Bell inequality. the reason is exactly the same.
One  cannot observe perfect correlations in the lab. Some noise is inevitable.
Therefore one must use Bell inequalities  of a new type (for the  first ones  see Mermin, 1990).
In the standard approach to multi-qubit Bell inequalities one assumes 
that each observer measures two randomly chosen dichotomic observables.
The results of measurements are used to compute the set of correlations
functions that one needs to check if there is a violation of certain 
Bell inequalities. 

A possible extension to this scenario is that 
each observer measures more than two local observables. Obviously
it cannot yield worse violation of local realism than in the standard
case. However, it is difficult to find analytically optimal Bell inequalities 
for more than two local observables and one must resort to numerical methods.

There is a computationally efficient method of finding the optimal
violation of local realism for arbitrary number of observers and
measured observables (\.Zukowski el al, 1999). It is based on well known linear optimisation 
algorithms. The greatest advantage of this method is that it gives 
necessary and sufficient conditions for the existence of local realistic 
description. The method has been successfully applied to the
problem of violation of local realism for entangled pairs of q-Nits 
($N=2,3,\dots,16$) and the numerical results of Kaszlikowski et al (2000) have been later confirmed 
analytically Kaszlikowski et al (2001), Collins et al (2001).

In this paper we show an application of the mentioned numerical method to
the GHZ correlations, i.e., to the maximally entangled state of three 
qubits. 

\section{Description of the method}
Let us consider the GHZ state of
three qubits 
\begin{eqnarray}
&&|\psi\rangle={1\over\sqrt 2}(|0\rangle_{1}|0\rangle_{2}|0\rangle_{3}+
|1\rangle_{1}|1\rangle_{2}|1\rangle_{3})
\end{eqnarray}
where $|i\rangle_j$ is the $i$-th state of the $j$-th qubit. 

Each observer measures the dichotomic observable
$\vec{n}\cdot\vec{\sigma}$, where $n=a,b,c$ ($a$ for the first
observer, $b$ for the second one and $c$ for the third one), $\vec{n}$ 
is a unit vector
characterising the observable which is measured by observer $n$ and
$\vec{\sigma}$ is a vector the components of which are standard Pauli
matrices. This family of observables 
$\vec{n}\cdot\vec{\sigma}$
covers all possible dichotomic observables for a qubit
system. 

The probability of obtaining
the result $m=\pm 1$ for the observer $a$, when measuring the
observable characterised by the vector
$\vec{a}$, the result $l=\pm 1$ for the observer $b$, when measuring the
observable characterised by the vector
$\vec{b}$ and the result $k=\pm 1$ for the observer $c$, when measuring the
observable characterised by the vector
$\vec{c}$ is equal to
\begin{eqnarray}
&P_{QM}(m,l,k;\vec{a},\vec{b},\vec{c})=
{1\over 8}(1+mla_3b_3+mka_3c_3+lkb_3c_3&\nonumber\\
&+mlk\sum_{r,p,s=1}^{3}M_{rps}a_r b_p c_s),&
\label{GHZpredictions}
\end{eqnarray}
where $a_r, b_p, c_s$ are components of vectors $\vec{a},\vec{b},\vec{c}$ 
and where nonzero elements of the tensor $M_{rps}$ are $M_{111}=1, M_{122}=-1, 
M_{212}=-1, M_{221}=-1$. In spherical coordinates vectors $\vec{a}, 
\vec{b}, \vec{c}$ read
\begin{eqnarray}
&&\vec{n}=(\cos\phi_n\sin\theta_n,\sin\phi_n\sin\theta_n,\cos\theta_n),
\label{GHZspherical}
\end{eqnarray}
where $0\leq \theta_n\leq \pi$ and $0\leq \phi_n \leq 2\pi$. From now on 
we will be
considering only the measurement of the observables characterised
by vectors with the zero third component, which is equivalent to
putting $\theta_{n}=\pi/2$. Thus, the formula (\ref{GHZpredictions})
acquires simpler form (we have replaced $\phi_a,\phi_b,\phi_c$ 
by $\alpha,\beta,\gamma$ respectively)
\begin{eqnarray}
&P_{QM}(m,l,k;\alpha,\beta,\gamma)=
{1\over 8}(1+mlk\sum_{r,p,s=1}^{3}M_{rps}a_r b_p c_s)&
\label{GHZpredictionsbis}
\end{eqnarray}
in which only the term responsible for three qubit correlations
is present. 

The probabilities of obtaining one of the results in the local
stations reveal no dependence on the local parameters,
$P_{QM}(l|\alpha)=P_{QM}(m|\beta)=P_{QM}(n|\gamma)
={1\over2}$. Similarly, the 
probabilities describing two qubit correlations do not
reveal dependence on the local parameters, i.e., 
$P_{QM}(l,m|\alpha,\beta)=P_{QM}(m,n|\beta,\gamma)=
P_{QM}(l,n|\alpha,\gamma)=
{1\over4}$. 

If there is a white noise in the quantum channel distributing qubits to
the observers we must replace the above quantum probabilities 
(\ref{GHZpredictionsbis}) by
\begin{eqnarray}
&P_{QM}^{V}(m,l,k|\alpha,\beta,\gamma)=
{1\over 8}(1+mlk V\sum_{r,p,s=1}^{3}M_{rps}a_r b_p c_s),&
\label{GHZpredictions2}
\end{eqnarray}
where $1-V$ ($0\leq V\leq 1$) is the amount of noise in the channel. The parameter $V$
is often called ``visibility''. 

Let us define the correlation function 
$E_{QM}^{V}(\alpha, \beta,\gamma)$ as 
$$
E_{QM}^{V}(\alpha, \beta,\gamma)=
\sum_{m,l,k=-1}^{1} mlkP(m,l,k;\alpha,\beta,\gamma)
V\sum_{r,p,s=1}^{3}M_{rps}a_r b_p c_s.$$
It equals $V\cos(\alpha+\beta+\gamma)$.  
As we see in the considered experiment there is no single and two qubit interference
and the correlation function, which depends only on three-particle correlations, contains all 
information about correlations in the system.  

In the experiment observer  
$a$ chooses between $N_a$ settings of the measuring 
apparatus $\alpha_1,\dots, \alpha_{N_a}$, observer
$b$ between $N_b$ settings $\beta_1,\dots, \beta_{N_{b}}$ 
and ,finally, observer
$c$ between $N_c$ settings $\gamma_1,\dots, \gamma_{N_{c}}$. 
For each triple of local settings we calculate
the quantum correlation function  $E_{QM}^{V}(\alpha_i,
\beta_j,\gamma_k)$, where $i=1,\dots,N_a,
j=1,\dots, N_b, k=1,\dots,N_c$. Thus we have a ``tensor'' 
$Q_{ijk}(V)=E_{QM}^{V}(\alpha_i,
\beta_j,\gamma_k)$ of quantum predictions. 

Within the local hidden variables formalism the correlation 
function must have
the following structure
\begin{equation} 
E_{LHV}(\alpha_i, \beta_j, \gamma_k)= 
\int d\rho(\lambda)A(\alpha_i,\lambda)B(\beta_j,\lambda)
C(\gamma_k,\lambda), 
\end{equation}
where for dichotomic measurements 
\begin{eqnarray}
&&A(\alpha_i,\lambda)=\pm1\nonumber\\
&&B(\beta_j,\lambda)=\pm1\nonumber\\  
&&C(\gamma_k,\lambda)=\pm1,
\end{eqnarray}
and they represent the values of local
measurements predetermined by the local hidden variables, 
denoted by $\lambda$, for the
specified local settings. This expression is an average over a certain
LHV distribution $\rho(\lambda)$ of certain {\it factorisable} ``tensors'', 
namely those with elements given by
$T_{ijk}(\lambda)=A(\alpha_i,\lambda)B(\beta_j,\lambda)
C(\gamma_k,\lambda)$.
Since the
only possible values of $A(\alpha_i,\lambda)$,
$B(\beta_j,\lambda)$ and $C(\gamma_k,\lambda)$ are $\pm1$ there are 
only $2^{N_{a}}$ {\it different} sequences of the values of $(A(\alpha_1,\lambda),\dots, A(\alpha_{N_{a}},\lambda))$, $2^{N_{b}}$ 
different sequences of 
$(B(\beta_1,\lambda),\dots,B(\beta_{N_{b}},\lambda))$,
$2^{N_{c}}$ 
different sequences of 
$(C(\gamma_1,\lambda),\dots, C(\gamma_{N_{c}},\lambda))$
and consequently they form only $2^{N_{a}+N_{b}+N_{c}}$ 
tensors $T_{ijk}(\lambda)$.

Therefore the structure of LHV models of $E_{LHV}(\alpha_i,
\beta_j,\gamma_k)$ reduces to discrete probabilistic models involving the
average of all the $2^{N_{a}+N_{b}+N_{c}}$ tensors $T_{ijk}(\lambda)$.  
In other
words, the local hidden variables can be replaced, without any loss of generality, by a
certain triple of variables $l,m,n$ that have integer values
respectively from $1,\dots,2^{N_{a}},1,\dots,2^{N_{b}},1,\dots,
2^{N_{c}}$.  To each $l$
we ascribe one possible sequence of the possible values of
$A(\alpha_i,\lambda)$, denoted from now on by $A(\alpha_i,l)$,
similarly we replace $B(\beta_j,\lambda)$ by $B(\beta_j,m)$ and
$C(\gamma_k,\lambda)$ by $C(\gamma_k,n)$ .
With this notation the possible LHV models of the correlation function
$E_{LHV}(\alpha_i, \beta_j, \gamma_k)$ acquire the following simple form
\begin{equation} 
  E_{LHV}(\alpha_i, \beta_j,\gamma_k)=
  \sum_{l=1}^{2^{N_{a}}}\sum_{m=1}^{2^{N_{b}}}\sum_{n=1}^{2^{N_{c}}}p_{lmn}
  A(\alpha_i,l)B(\beta_j,m)C(\gamma_k,n),
  \label{MODELGHZ}
\end{equation}
with, of course, the probabilities satisfying $p_{lmn}\geq0$ and
$$\sum_{l=1}^{2^{N_{a}}}\sum_{m=1}^{2^{N_{b}}}
\sum_{n=1}^{2^{N_{c}}}p_{lmn}=1$$. 
Please note that not all tensors $T_{ijk}(lmn)$
are different (in fact only half of them differ). 

The conditions for local hidden variables to reproduce the quantum prediction with a
visibility $V$ can be simplified to the problem of maximising a
parameter $V$ for which exists a set of $2^{N_a+N_b+N_c-1}$ probabilities 
$\tilde{p}_{lmn}$, such that 
\begin{equation}
  \sum_{l=1}^{2^{N_a}}\sum_{m=1}^{2^{N_b}}\sum_{n=1}^{2^{N_c-1}}
  \tilde{p}_{lmn}A(\vec{a}_i,l)B(\vec{b}_j,m)C(\vec{b}_k,n)
  =Q_{ijk}(V).
  \label{MODEL1}
\end{equation}
Because, for the given local settings (\ref{MODEL1}) 
imposes linear constraints on the probabilities and the
visibility, and we are looking for the maximal $V$, the problem
can be solved by means of linear programming methods of optimisation.

We want to find such local
settings for which this maximal $V$ reaches its
minimum. This is due to the fact that in such a case the noise admixture, $1-V$,
is maximal. in such a case the non-classical matrix of quantum correlations reveals the strongest 
resistance with respect to white noise admixtures. This can be treated as a measure of
the ``strength'' of violation of local realism. 

The set of linear equations (\ref{MODEL1}) constitute a certain region
in a $D=2^{N_a+N_b+N_c-1}+1$ dimensional real space- $2^{N_a+N_b+N_c-1}$ probabilities
plus the visibility. The border of the region
consists of hyper planes each defined by one of the equations belonging
to (\ref{MODEL1}); thus, if the equations do not contradict each other 
the region is a convex set with a certain number of vertices. 
On this convex set we define a linear function (cost function)
$f(p_1,\cdots,p_{2^{N_a+N_b+N_c-1}},V)=V$, and we seek its  maximum.

The fundamental
theorem of linear programming states that {\it the cost function 
reaches its maximum at one of the vertices}. Hence, it suffices to
find numerical values of the cost function calculated at the vertices and
then pick up the largest one. Of course, the algorithmic implementation
of this simple idea is not so easy, for we must have a method of finding
the vertices, for which the value of the function continually 
increases, so 
that the program
reaches the optimal solution in the least possible number of steps. 
Calculating the value of the cost function
at every vertex would take too much time, as there may be too many
of them.

There are lots of excellent algorithms which solve the above optimisation
problem. Here we have used the algorithm invented by Gondzio (1995)
and implemented in the commercial code HOPDM 2.30 (Higher Order Primal-Dual
Method) written in C programming language.

However, finding the maximal visibility for the given local settings
of the measuring apparatus is not enough. We should remember that our main goal
is to find such local setting for which the threshold visibility is the lowest
one. The maximal visibility $V^{max}$ returned by the HOPDM 2.30 procedure 
depends on the local settings entering right hand side of (\ref{MODEL1}). 
Thus, returned $V^{max}$ 
can be treated as the many variable function, which depends on $N_a+N_b+N_c$
angles in the coplanar case and two times more in the non coplanar one, i.e.,
$V^{max}=V^{max}(\vec{a}_1,\dots,\vec{a}_{N_a},\vec{b}_1,\dots,\vec{b}_{N_b})$.

Hence, we should also have a numerical procedure which finds the minimum
of $V^{max}$. 
Because we do not know much about the structure of $V^{max}$ as a function
of the local settings
the only reasonable
method of finding the $V^{max}$ minimum is the Downhill Simplex Method (DSM) (Nelder and Mead,  1965). 
The way it works toward finding the extremum is the following. If the 
dimension of the domain of a function is $Dim$ the DSM randomly generates
$Dim+1$ points. This way it creates a starting simplex with vertices being the points. 
these points. Then it calculates the value of a function at the vertices and
starts exploring the space by stretching and contracting the simplex. In every
step if it 
finds a vertex where the value of the function is lower then in others it 
"goes" in this direction.

We have checked 
four cases: $N_{a}=N_{b}=N_{c}=2,3,4,5$ with
the result that the threshold visibility admitting local hidden variable model 
is $V={1\over 2}$.
This result is in concurrence with 
the threshold visibility obtained earlier in Mermin (1990)
with the usage of appropriate Bell inequalities.

Because of the complexity of the space being the domain
of the $V^{max}$ function to find a global minimum for the case $N_{a}=N_{b}
=N_{c}=2,3$ we 
have run the amoeba procedure
30 times with varied starting points. For $N_{a}=N_{b}=N_{c}=4$ 
we calculated $V^{max}$
on $9000$ randomly chosen sets of the local settings whereas in the
case $N_{a}=N_{b}=N_{c}=5$ we have calculated $V^{max}$ on the following
set of the local settings: $\alpha_1=0,\alpha_2=\pi/8,
\alpha_3=\pi/4,\alpha_4=3\pi/8,\alpha_5=\pi/2,\beta_1=\gamma_1=-\pi/4,
\beta_2=\gamma_2=-\pi/8,\beta_3=\gamma_3=0,\beta_4=\gamma_4=\pi/8,
\beta_5=\gamma_5=\pi/4$. In both cases the only reason for abandoning the DSM method
was the exploding computational time.

An interesting feature of the results is that, the 
threshold visibility
$V^{max}={1\over2}$ is always achieved for such 
settings of the measuring
apparatus which include as a subset the settings 
giving maximal violation of the inequalities. Similar result was  obtained for two maximally  
entangled qubit,
also  using the  numerical  method,  by \.Zukowski et al (1999) and Massar  et al (2002). 

\section{Conclusions}
The presented numerical approach to the three qubit GHZ correlations
gives the sufficient and necessary conditions for the existence
of local hidden variables for the given experimental situation, i.e., 
for the fixed number of positions of the measuring apparatus at each side 
of the experiment. 

For the cases of $N_a=N_b=N_c=2,3$ we have found such numerical
values of the local settings for which the critical visibility admitting local hidden variables
has the lowest possible value. Up to the possibility that the DSM 
procedure has not succeeded in finding the global minimum of
$V^{max}$ the visibility $V={1\over2}$ is the ultimate limit drawing
the borderline between local hidden variables and quantum mechanics for these cases, i.e., for
2 and 3 settings of the measuring apparatus at each side of the experiment. 

For $N_a=N_b=N_c=4$ the critical visibility returned by the program for every
random choice of local settings has been always higher 
than $\frac{1}{2}$.

In the last case, i.e., for $N_a=N_b=N_c=5$, we have found
the threshold value for local settings including as a subset
settings giving maximal violation of a three particle Bell-type inequality 
with the result again $V^{max}={1\over 2}$ (the DSM
has not been used).

Unfortunately, due to the computer time and memory limitations we
could not check more settings of the measuring apparatus. Nevertheless, one
could  possibly conjecture
that increasing the number of settings will not lead to a critical visibility
lower than $V={1\over 2}$. This quite surprising especially when  one considers  the  
fact that already three  settings per observation site lower the  critical  visibility
for four GHZ  qubits  or more (\.Zukowski and  Kaszlikowski, 1997).

The important aspect of the presented analysis of the GHZ correlations
is that the numerical approach can be directly applied to measurement data.

DK acknowledges NUS Grant WBS: R-144-000-089-112.
MZ acknowledges KBN Grant PBZ KBN 043/P03/2001.

\section*{References}

\noindent
A. Aspect, J. Dalibard  and G. Roger, 1982, Phys. Rev. Lett. {\bf 47}, 460.

\noindent
J. S. Bell, 1964, Physics,  {\bf 1}, 195.

\noindent
J. S. Bell, 1966, Rev. Mod. Phys., {\bf 38}, 447.

\noindent
A. Einstein, B. Podolsky and N. Rosen, 1935, Phys. Rev., {\bf 47}, 777.

\noindent
J. F. Clauser and M. A.  Horne, 1974, Phys. Rev. D {\bf 10}, 526

\noindent
J. F. Clauser, M. A.  Horne,  A.  Shimony and R. A.  Holt, 1969, Phys. Rev. Lett., {\bf 23}, 880.

\noindent
 D. Collins, N. Gisin, N. Linden, S. Massar, and S. Popescu, 2002,  Phys. Rev. Lett. {\bf 88},
040404 (2002).

\noindent
A. K. Ekert, 1991. Phys. Rev. Lett., {\bf 67}, 661.

\noindent
J. S. Freedman and J. F. Clauser, 1972, Phys. Rev. Lett., {\bf 28} 938

\noindent
 J. Gondzio, 1995,  European Journal of Operational
Research {\bf 85}, 221.

\noindent
D. M. Greenberger, M. A. Horne and A. Zeilinger, 1989,
in {\it Bell's Theorem, Quantum Theory, and Conceptions of the Universe,}
edited by Kafatos,\ M. (Kluwer Academics, Dordrecht, The Netherlands), 
p. 73.

\noindent
D. Kaszlikowski, P. Gnacinski, M. \.Zukowski, W. Miklaszewski, and A. Zeilinger, 2001, Phys. Rev. Lett.
{\bf 85}, 4418.

\noindent
 D. Kaszlikowski, L. C. Kwek, Jing-Ling Chen, M.  \.Zukowski, and C. H. Oh
Phys. Rev. A 65, 032118 (2002).

\noindent
S. Massar, S. Pironio, J. Roland and B. Gisin, 2002,  Phys. Rev. A {\bf 66}, 052112.

\noindent
N. D. Mermin, 1990, Phys. Rev. Lett. {\bf 65}, 1838. 

\noindent
J. A. Nelder and R. Mead, 1965, Computer Journal
{\bf 7}, 308.

\noindent
I. Pitowsky and K. Svozil, 2001,
     Phys. Rev. A { \bf64}, 014102.

\noindent
H. Weinfurter, M. \.Zukowski, 2001, Phys. Rev. A, {\bf 64}, 010102R.

\noindent
R. F. Werner and M. M. Wolf, 2001,  Phys. Rev. A {\bf 64}, 032112.

\noindent 
M. \.Zukowski and C. Brukner, 2002, Phys. Rev. Lett. 88, 210401

\noindent
M. \.Zukowski, D. Kaszlikowski, A. Baturo and J. -A. Larsson, 1999, e-print  quant-ph/9910058.

\noindent
M. \.Zukowski and D. Kaszlikowski, 1997, Phys. Rev. A {\bf 56}, R1685.

\end{document}